\documentclass{emulateapj}
\usepackage{natbib}
\usepackage{apjfonts}
\usepackage{multirow}

\newcommand{\msigma}{$M_{\rm BH}-\sigma_{\star}$}
\newcommand{\ml}{$M_{\rm BH}-L_{\rm bulge}$}

\newcommand{\actaa}{Acta Astronomica}
\newcommand{\hst}{{\it HST}}

\shorttitle{A Cepheid Distance to NGC\,6814}
\shortauthors{Bentz et al.}

\received{}
\accepted{}

\begin{document}

\title{A Cepheid-based Distance to the Seyfert Galaxy NGC\,6814}

\author{ Misty~C.~Bentz\altaffilmark{1},
Laura~Ferrarese\altaffilmark{2,3},
Christopher~A.~Onken\altaffilmark{4,5},\\
Bradley~M.~Peterson\altaffilmark{6,7,8},
Monica~Valluri\altaffilmark{9},
}

\altaffiltext{1}{Department of Physics and Astronomy,
		 Georgia State University,
		 Atlanta, GA 30303, USA;
		 bentz@astro.gsu.edu}
\altaffiltext{2}{NRC-Herzberg Astronomy and Astrophysics, 
    5071 West Saanich Road, 
    Victoria, BC, V9E 2E7, Canada}		 
\altaffiltext{3}{Gemini Observatory, 
        Northern Operations Center, 
        670 N.\ A'ohoku Place, Hilo, HI 96720, USA}
\altaffiltext{4}{Research School of Astronomy and Astrophysics, 
        Australian National University, 
        Canberra, ACT 2611, Australia }
\altaffiltext{5}{ARC Centre of Excellence for All-sky Astrophysics (CAASTRO)}
\altaffiltext{6}{Department of Astronomy, The Ohio State University, 140 W 18th Ave, Columbus, OH 43210, USA}
\altaffiltext{7}{Center for Cosmology and AstroParticle Physics, The Ohio State University, 191 West Woodruff Ave, Columbus, OH 43210, USA}
\altaffiltext{8}{Space Telescope Science Institute, 3700 San Martin Drive, Baltimore, MD 21218}
\altaffiltext{9}{Department of Astronomy,
         University of Michigan,
         Ann Arbor, MI, 48104, USA}

\begin{abstract}
We present a Cepheid-based distance to the nearby Seyfert galaxy NGC\,6814 from {\it Hubble Space Telescope} observations.  We obtained F555W and F814W imaging over the course of 12 visits with logarithmic time spacing in 2013 August$-$October.  We detected and made photometric measurements for 16,469 unique sources across all images in both filters, from which we identify 90 excellent Cepheid candidates spanning a range of periods of $13-84$\,days.  We find evidence for incompleteness in the detection of candidates at periods <21\,days. Based on the analysis of Cepheid candidates above the incompleteness limit, we determine a distance modulus for NGC\,6814 relative to the LMC of $\mu_{\rm rel\,LMC}=13.200^{+0.031}_{-0.031}$\,mag. Adopting the recent constraint of the distance modulus to the LMC determined by \citeauthor{pietrzynski19}, we find $m-M=31.677^{+0.041}_{-0.041}$ which gives a distance of $21.65 \pm 0.41$\,Mpc to NGC\,6814.  
\end{abstract}

\keywords{galaxies: active --- galaxies: nuclei --- galaxies: Seyfert}

\section{Introduction}

The Leavitt Law, describing the relationship between the luminosity of Cepheid stars and their periods of variability \citep{leavitt1912}, is one of the most revolutionary discoveries in the field of astronomy.  The application of the Leavitt Law to a single variable star in the Andromeda Galaxy by \citet{hubble1925} settled the 1920 Great Debate and led to the understanding that the universe was far larger than our Milky Way galaxy.  Nearly a century later, Cepheid variable stars continue to be a key tool for exploring our location relative to other galaxies, and for constraining the age and evolution of the universe.  An accurate measurement of the expansion rate of the universe based on observations of extragalactic Cepheids was one of the key motivations behind the construction of the {\it Hubble Space Telescope} (\hst) and one of its original Key Projects \citep{kennicutt95,freedman01}.

But \hst\ also led to several surprises.  One completely unexpected result was the discovery that every massive galaxy is likely hosting a supermassive black hole in its nucleus.  Furthermore, when the masses of the black holes are constrained, they are found to correlate with other properties of the host galaxies on much larger spatial scales, such as the bulge velocity dispersion (\msigma, \citealt{ferrarese00,gebhardt00,gultekin09}) and the bulge luminosity (\ml, \citealt{kormendy95,kormendy13}).

Reverberation mapping \citep{blandford82,peterson93} is one of the few techniques that can be used to directly constrain the mass of a black hole.  Reverberation mapping relies on light echoes in the photoionized gas around an active black hole to constrain the size of the region.  Combining the time delay of the echo with the Doppler-broadened width of a broad emission line provides a direct constraint on the black hole mass.  While this technique can only be applied to broad-lined active galactic nuclei (AGNs), it is distance-independent, in principle, it may be applied at any redshift because it relies on time resolution rather than spatial resolution.

In the very local universe, bright broad-lined AGNs are rare.  Most of what we know about nearby galaxies and black holes (e.g., \msigma, \ml, and other scaling relationships) is instead derived from quiescent galaxies.  In these cases, dynamical modeling of the spatially resolved bulk motions of stars or gas are used to directly constrain the black hole mass (cf.\ the review by \citealt{kormendy13}).  Stellar dynamical modeling is generally considered to be more robust than gas dynamical modeling because of its immunity to turbulence and other non-gravitational effects.  The reliance on spatial resolution, however, makes all dynamical mass measurements fundamentally dependent on the assumed distance.

\begin{figure*}
    \centering
    \epsscale{0.8}
    \plotone{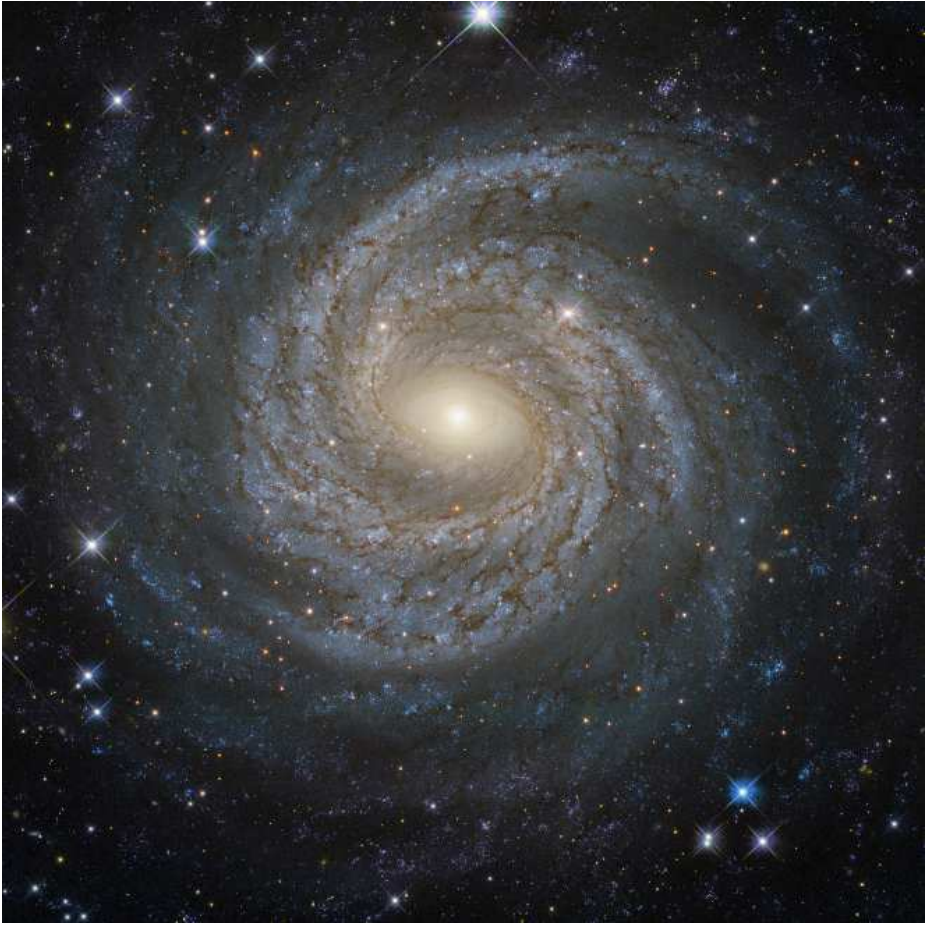}
    \caption{ \hst\ Wide Field Camera 3 composite image of NGC\,6814 through the F555W, F814W, and F160W filters. North is oriented $51^{\circ}$ counterclockwise from the top and the image is $2\farcm6 \times 2\farcm6$.  Image credit: Judy Schmidt.}
    \label{fig:n6814hst}
\end{figure*}

Observational and computational efforts to understand the co-evolution of black holes and galaxies across cosmic time thus rely on the results of reverberation mapping to constrain black hole masses at moderate- to high-redshift, and the results of dynamical modeling to constrain black hole masses at low redshift.  The different technical requirements between these methods have resulted in very few attempts to compare the black hole masses from multiple techniques in the same galaxies.  To date, the only published comparisons between reverberation mapping and stellar dynamical modeling are for NGC\,4151 \citep{bentz06b,onken14,derosa18} and NGC\,3227 \citep{davies06,denney10,derosa18}.  The masses generally agree, although in the case of NGC\,3227 the dynamical mass is larger by $\sim 0.5$\,dex.

We have thus begun a program to increase the number of direct black hole mass comparisons between reverberation mapping and stellar dynamical modeling.  The rarity of nearby broad-lined AGNs is the main limitation to our efforts, but there are a handful of excellent targets that meet the technical requirements for both techniques, including NGC\,6814.   

NGC\,6814 is a nearly face-on grand design spiral (see Figure~\ref{fig:n6814hst}) and is also one of the nearest ($z=0.0054$) broad-lined Seyfert galaxies.  The central supermassive black hole has a reverberation-based mass measurement of $(1.85 \pm 0.35) \times10^7$\,$M_{\odot}$ \citep{bentz09c}.  We successfully obtained spatially resolved measurements of the nuclear stellar kinematics with the Gemini-North Near-infrared Integral Field Spectrograph (NIFS), which we will present in a future paper.  The timing arguments that underlie reverberation-based mass measurements are independent of the distance, but the stellar dynamical mass is linearly dependent on the assumed distance to the galaxy.  NGC\,6814 is too nearby to assume that it is in the Hubble flow, so its redshift cannot be relied upon to provide an accurate distance estimate.  Furthermore, its orientation, being nearly face on, obviates the possibility of a distance estimate based on the galaxy luminosity and the width of the \ion{H}{1} 21\,cm emission line \citep{tully77}.

Accordingly, we undertook a program to monitor NGC\,6814 with \hst\ imaging so that we could detect and characterize Cepheid variable stars in the galaxy.  The Cepheid candidates that we describe below, combined with the most recent calibration of the Leavitt Law, provide the first accurate distance to NGC\,6814.

\section{Observations}

Observations of NGC\,6814 were conducted with the {\it Hubble Space Telescope} Wide Field Camera 3 (WFC3) between 2013 August 08 and 2013 October 09.  The UVIS detectors  provide images with a field of view of $2\farcm6 \times 2\farcm6$ and pixel scale of $0\farcs04$.

Over the course of 12 visits, with logarithmic time spacing between them, we devoted a single orbit to imaging the galaxy through the F555W filter.  At shorter wavelengths, Cepheids are known to show a larger amplitude of variations but they are also more severely affected by reddening, a concern for NGC\,6814 given its close proximity to the Galactic plane ($b=-16.0^{\circ}$ with $A_B=0.67$\,mag, based on the \citealt{schlafly11} recalibration of the \citealt{schlegel98} dust map). Thus, the F555W band is a reasonable compromise between maximizing the amplitude of variations and minimizing the extinction and reddening caused by dust \citep{freedman94}. The available exposure time in each orbit was split equally among four images following a standard 4-point dither pattern.  On five of the 12 visits, a second orbit was dedicated to observations through the F814W filter, and the available time was also split equally into four exposures with a standard 4-point dither pattern.  
Table~\ref{tab:obs} provides a summary of the observations.  All of the images were obtained with a fixed orientation to facilitate their registration across all visits and filters.

The final analysis was carried out on images that were downloaded from MAST after 2016 February, and thus were corrected for charge transfer efficiency losses.  We first drizzled all of the F555W images together with the AstroDrizzle pipeline to create a deep F555W image of the galaxy.  This deep image was then used as the reference for drizzling each individual visit and each filter onto a common grid.  The effective exposure time for each drizzled image was 2400\,s in F555W and 2520\,s in F814W,
which corresponds to an estimated signal-to-noise ratio S/N$\approx 10$ for a Cepheid variable star with a period of 20\,days at a distance of 22\,Mpc (similar to the distance predicted by the redshift of NGC\,6814).

\begin{deluxetable}{lcccc}
\tablecolumns{8}
\tablewidth{0pt}
\tablecaption{Observations}
\tablehead{
\colhead{Visit} &
\colhead{$UT$} &
\colhead{filter} &
\colhead{exp.\ time} &
\colhead{dataset} 
\\
\colhead{} &
\colhead{(yyyy-mm-dd)} &
\colhead{} &
\colhead{($s$)} &
\colhead{} 
}
\startdata
1  & 2013-08-10 & F555W   & 2400.0  & IBY101010   \\
   &            & F814W   & 2520.0  & IBY101020   \\
2  & 2013-08-21 & F555W   & 2400.0  & IBY102010   \\
3  & 2013-08-27 & F555W   & 2400.0  & IBY103010   \\
   &            & F814W   & 2520.0  & IBY103020   \\
4  & 2013-08-31 & F555W   & 2400.0  & IBY104010   \\
5  & 2013-09-05 & F555W   & 2400.0  & IBY105010   \\
6  & 2013-09-08 & F555W   & 2400.0  & IBY106010   \\
7  & 2013-09-10 & F555W   & 2400.0  & IBY107010   \\
   &            & F814W   & 2520.0  & IBY107020   \\
8  & 2013-09-11 & F555W   & 2400.0  & IBY108010   \\
9  & 2013-09-19 & F555W   & 2400.0  & IBY109010   \\
10 & 2013-09-25 & F555W   & 2400.0  & IBY110010   \\
   &            & F814W   & 2520.0  & IBY110020   \\
11 & 2013-10-02 & F555W   & 2400.0  & IBY111010   \\
12 & 2013-10-09 & F555W   & 2400.0  & IBY112010   \\
   &            & F814W   & 2520.0  & IBY112020   
%
\label{tab:obs}
\enddata 

\end{deluxetable}

\begin{figure*}
    \centering
    \epsscale{0.8}
    \plotone{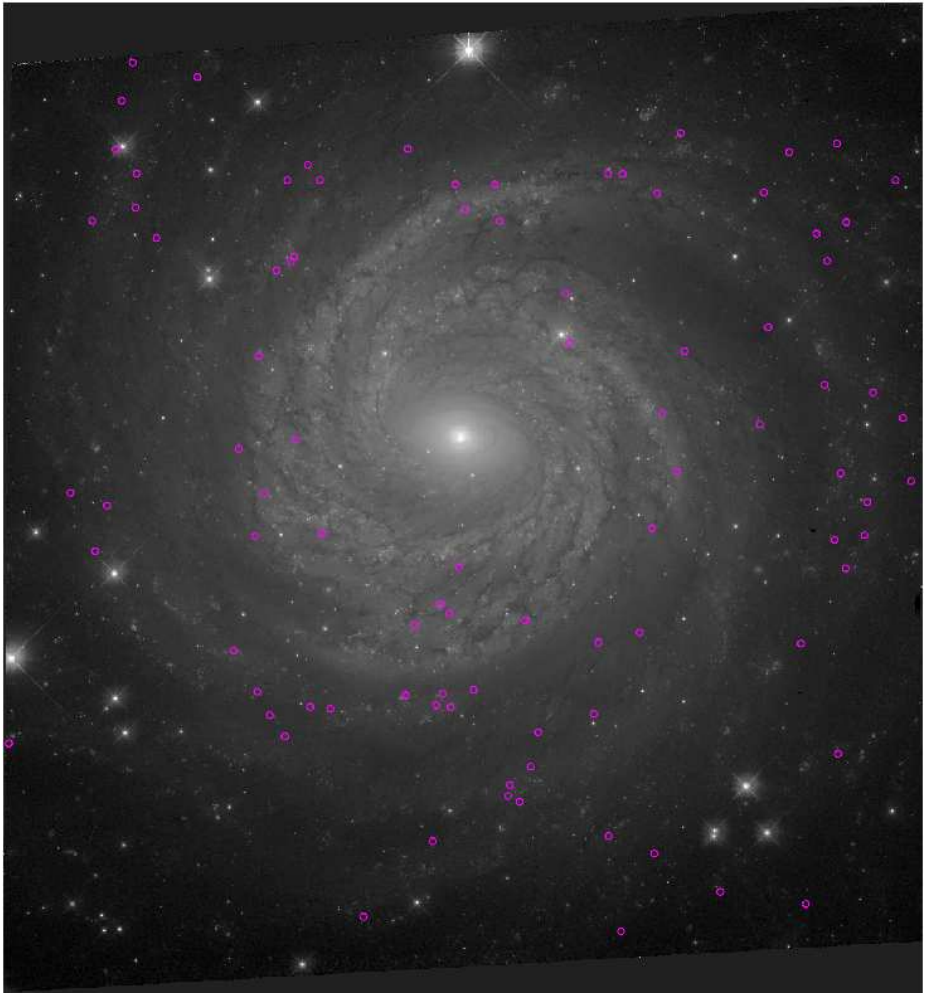}
    \caption{Deep F555W image of NGC\,6814 with the locations of the final sample of candidate Cepheid stars marked by magenta circles.}
    \label{fig:n6814cands}
\end{figure*}

\section{Photometry}

To begin searching for Cepheid variable stars in NGC\,6814, we followed the standard procedures of identifying and carrying out PSF photometry for all point sources in the frames using {\tt DAOPHOT}, {\tt ALLSTAR}, and {\tt ALLFRAME} \citep{stetson87,stetson94}.

Point sources were first identified in the deep F555W frame, and then photometric measurements were made for every object in the source list in the drizzled frames for each individual visit.  Based on a sample of 131 relatively bright and uncrowded stars, we determined the differences between the fitted PSF magnitudes and the magnitude within a standard WFC3 aperture radius of 10\,pixels, thus providing aperture corrections in F555W and F814W.

We then matched the source lists between each of the individual visits and filters.  The final source list includes only those objects found across all 12 visits in F555W and all five visits in F814W, leaving us with 16,469 unique point sources in the field of NGC\,6814.

The photometry was then calibrated by adopting the zero-points for a 10\,pixel aperture radius as used by \citet{riess19}: 25.727\,mag for F555W and 24.581\,mag for F814W in the vegamag system.

\begin{figure*}
\centering
\epsscale{1.1}
\plotone{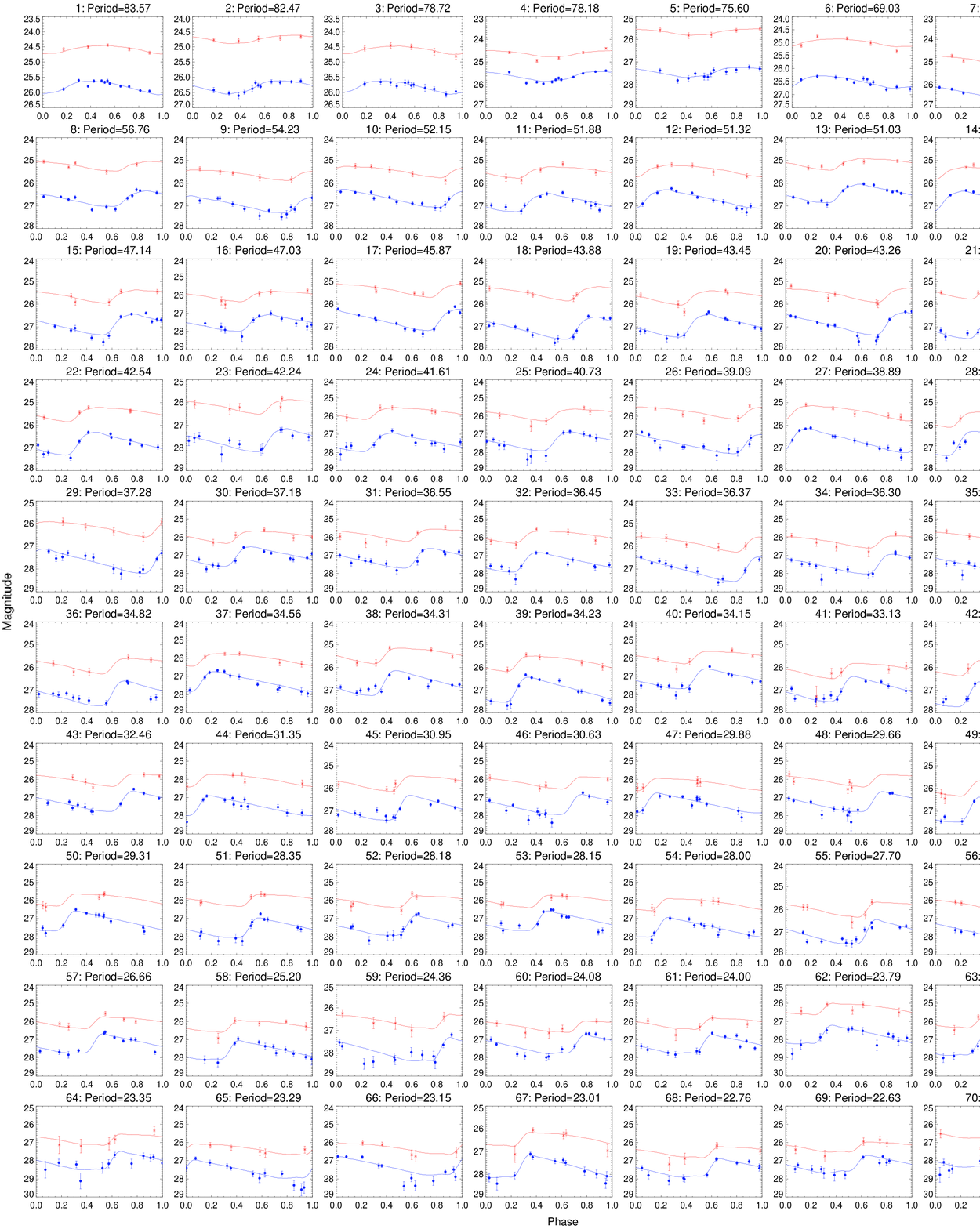}
\caption{F814W ({\it red}) and F555W ({\it blue}) light curves of Cepheid candidates
  with the LMC Cepheid template fits overlaid as the solid curves.}
\label{fig:lcs}
\end{figure*}

\begin{figure*}
\centering
\epsscale{1.1}
\plotone{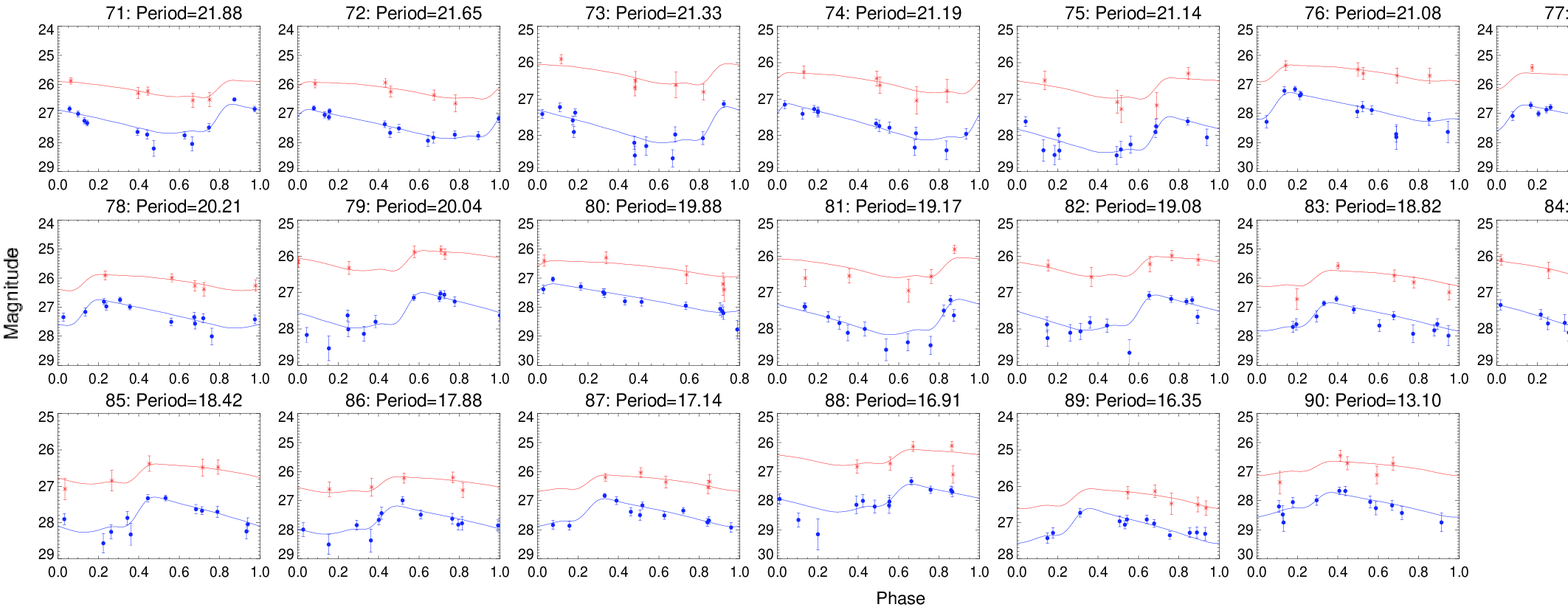}
\caption{Figure~\ref{fig:lcs} continued.}
\end{figure*}

\section{Cepheid Candidate Selection}

For each point source detected across all of the frames, we calculated the \citet{welch93} variability index as:
\begin{equation}
I_V = \sqrt{ \frac{1}{n(n-1)}}\sum_{i=1}^{n} \delta V_i \delta I_i  
\end{equation}
where $\delta V_i$ and $\delta I_i$ are the magnitude residuals in F555W and F814W, respectively, relative to the mean and normalized by the uncertainty. We selected those objects with $I_V > 1.25$ as the most likely sources showing true variability, which left us with 826 objects. 

We then fit the F555W and F814W light curves of all of these objects with the Cepheid templates of \citet{yoachim09}, focusing on the LMC long period ($P>10$\,days) subsample.  We began with an initial period search that covered the range $10\leq P \leq 100$\,days with 150 steps equally spaced in $\log P$.  For each object, we retained the six iterations with the lowest $\chi^2$ and compared the best-fit periods. For closely spaced steps in the initial guess for the period and well-behaved light curves with real Cepheid features, the best-fit period should remain fairly constant over a moderate range of initial guesses.  We thus retained only the candidates with $\Delta P < 0.3$\,days between the six iterations with the lowest $\chi^2$, leaving us with 497 candidates.  

We also calculated the reduced $\chi^2$ for the assumption of a straight line fit to the F555W light curves, retaining only those objects that were better fit by the Cepheid templates than a straight line.  Finally, we required the best-fit period to be within $10.0 \leq P \leq 85.0$\,days, and required the candidates to have best-fit colors in the range $0.5 \leq F555W-F814W \leq 2.0$\,mag, as expected for Cepheids in the instability strip, but with a more generous upper limit to account for the excess Galactic extinction along the line of sight to NGC\,6814.  This left us with 197 candidates.  Based on the recommendations of \citet{macri06} and \citet{shappee11}, we also verified that all candidates had amplitude ratios between 0.25 and 0.75\,mag in F555W, and uncertainties in their $F555W-F814W$ colors of $<0.3$\,mag.

The remaining 197 candidates were subjected to visual inspection, in which the photometry and the best-fit template light curves were examined for the characteristic Cepheid "sawtooth" shape.   Many of the candidates were spurious, likely the result of defects in the detector or cosmic rays.  Only 103 candidates passed the visual inspection with an obvious or likely sawtooth shape. For these 103 candidates, we recorded the period, phase, mean F555W magnitude, and mean F814W magnitude of the best template fits to the observed light curves.

Finally, the mean magnitudes were corrected for the effects of crowding in the images, as follows.  We conducted artificial star tests in which we randomly injected 100 artificial stars into small portions of the full frame centered on each of the Cepheid candidates.  Each artificial star was injected within a radius of 20\,pixels ($0\farcs8$) from the candidate, and with $-0.5 \leq \Delta m \leq 0.5$\,mag relative to the candidate.  We then identified and made photometric measurements of all the point sources in all of the subimages, in the same way as before.  The typical difference between the injected magnitude and the recovered magnitude was found to be $0.011 \pm 0.018$\,mag in F555W and $0.017 \pm 0.018$\,mag in F814W, with crowding effects biasing the measured magnitudes slightly brighter.

\begin{figure}
\epsscale{1.15}
\plotone{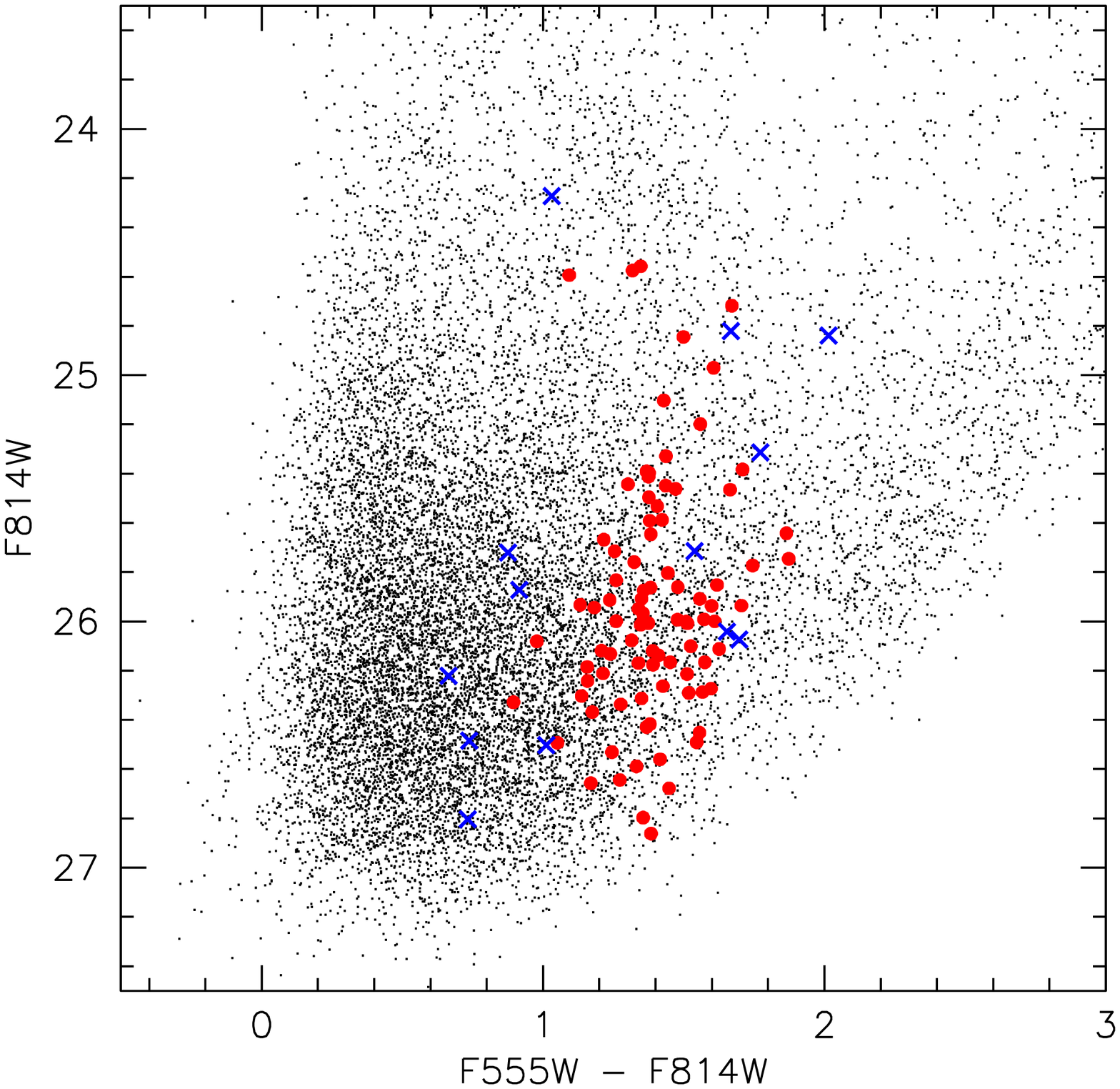}
\caption{Color-magnitude diagram of the 16,469 point sources detected in the field of NGC\,6814.  Cepheid candidates are highlighted in red, and candidates that were determined to be >3$\sigma$ outliers or blue blends are shown as blue crosses.}
\label{fig:cmd}
\end{figure}

\begin{figure}
\epsscale{1.15}
\plotone{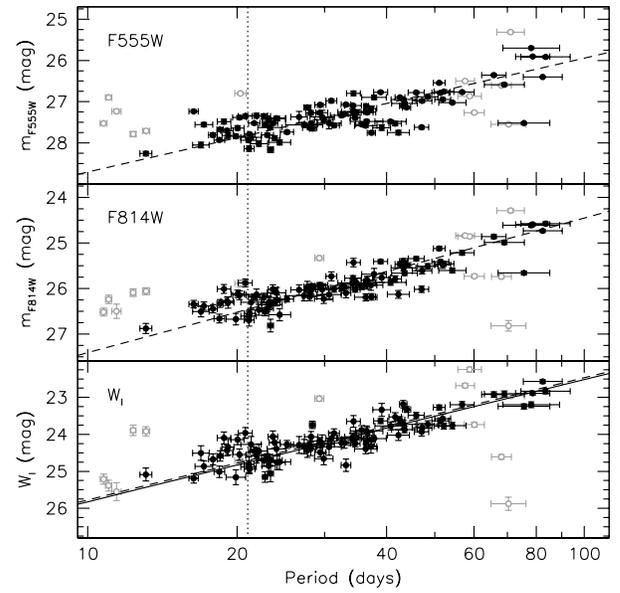}
\caption{Period-luminosity relationships for Cepheid candidates in F555W, F814W, and Wesenheit magnitudes.  The filled points are the 90 candidates that passed the outlier clipping and the open points are those that did not.  The dashed lines are the best fits to the points assuming the slopes found for fundamental mode classical Cepheids in the LMC by \citet{riess19} and including all the filled points.  The vertical dotted line marks $P=21$\,days, and the solid line in the bottom panel displays the best fit to the Wesenheit magnitudes for candidates with $P>21$\,days.}
\label{fig:pl}
\end{figure}

\begin{figure}
\epsscale{1.15}
\plotone{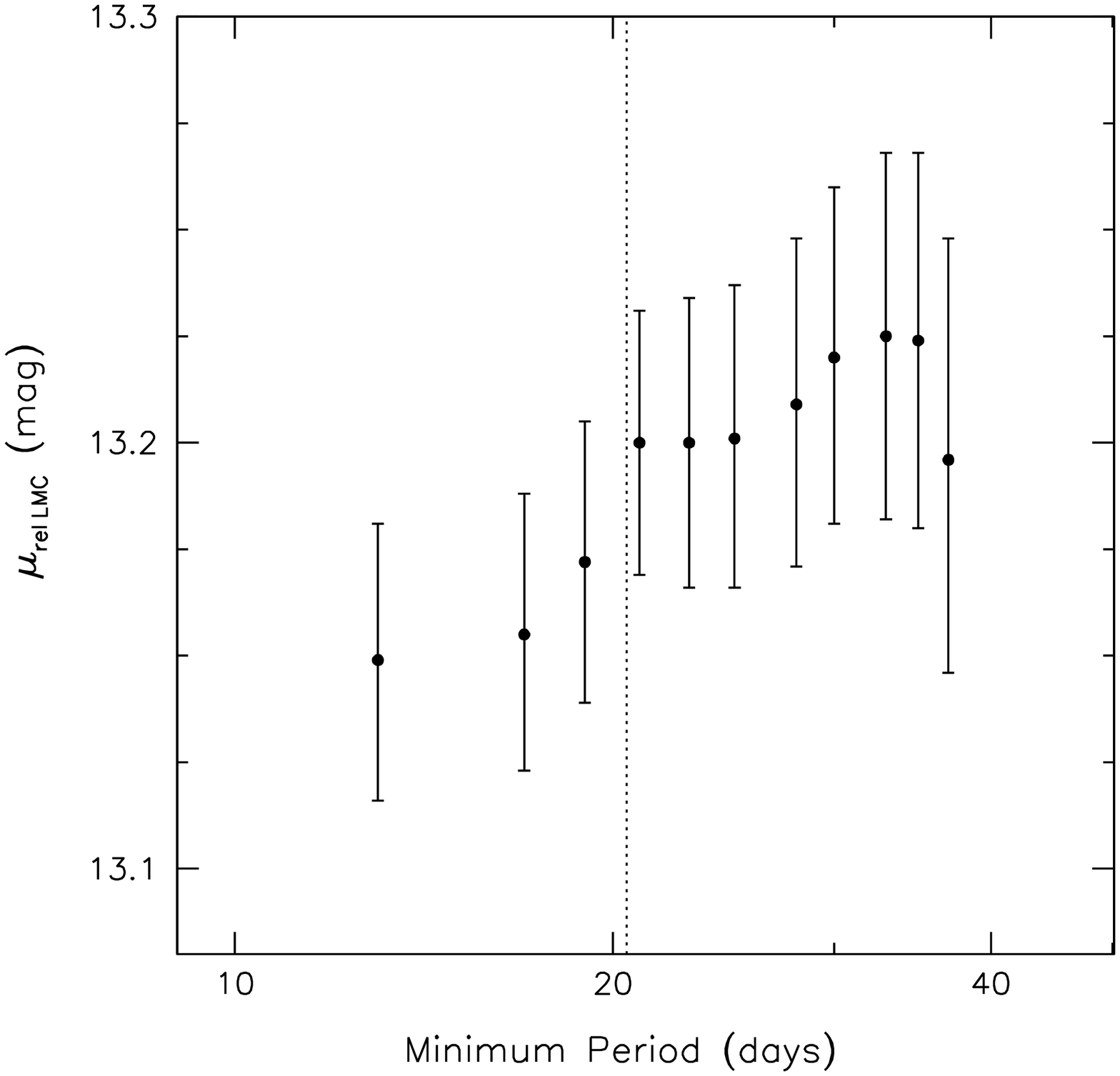}
\caption{Distance modulus relative to the LMC as a function of the minimum period of Cepheid candidates.  The vertical dotted line marks $P=21$\,days.  Below this threshold, there is clear evidence for incompleteness among the candidates.}
\label{fig:minp}
\end{figure}

\section{Period-Luminosity Relationship}

Based on the parameters constrained by the fits to the light curves in the previous section, we constructed Leavitt laws for the Cepheid candidates using F555W, F814W, and Wesenheit magnitudes \citep{madore82}, defined as:
\begin{equation}
    W_I = m_{F814W} - R(m_{F555W}-m_{F814W})
\end{equation}
where $R = A_I / (A_V -A_I)$.  Wesenheit magnitudes are less susceptible to reddening, both along the the line of sight and intrinsic to the galaxy.  We follow \citet{riess19} by adopting $R=1.3$ based on the \citet{cardelli89} extinction curve.

The calibration of LMC Cepheids with \hst\ F555W and F814W photometry by \citet{riess19} means that we can avoid any potential filter offsets or color biases when analyzing the Cepheids in NGC\,6814.  We adopt the \citet{riess19} calibrations of the period-luminosity relationship for F555W, F814W, and W$_I$ magnitudes reported in their Table~3.  Holding the slope ($\alpha$) fixed, we solved for the intercept ($\beta$) of the relationship as:
\begin{equation}
    m = \alpha \log P + \beta.
\end{equation}
Linear regressions were carried out with a Bayesian approach using the {\sc linmix\_err} algorithm \citep{kelly07}, which includes measurement errors in both coordinates and a component of intrinsic, random scatter.

Following the recommendation of \citet{kodric15} and the procedures of \citet{hoffmann16}, we then iteratively determined the best fit and rejected the most significant outlier in the $W_I$ relationship until every candidate was within $3\sigma$ of the best-fit period-Wesenheit magnitude relationship.  These outliers are likely to be true variable sources that are either misidentified as Cepheids or misclassified as fundamental single-mode pulsators. Finally, we also rejected any remaining candidates that were bright blue outliers, which we classified as $m_{\rm pred} - m_{\rm obs} \gtrsim 1.0$\,mag in the F555W filter, as these candidates are likely to be blends.  After the $3\sigma$ clipping and removal of blue outliers, we were left with 90 excellent Cepheid candidates with periods ranging from 13 to 84\,days (see Figure~\ref{fig:lcs}).  Their properties are listed in Table~\ref{tab:cands} and their positions within NGC\,6814 are denoted by the magenta circles in Figure~\ref{fig:n6814cands}.  In Figure~\ref{fig:cmd} we display a color-magnitude diagram of all the point sources detected in the field of NGC\,6814 with the candidates and outliers highlighted.

The best-fit period-luminosity relationships for F555W, F814W, and $W_I$ after clipping outliers are  displayed in Figure~\ref{fig:pl} and tabulated in Table~\ref{tab:pl}.  We also list in Table~\ref{tab:pl} the distance modulus of NGC\,6814 relative to the LMC, $\mu_{\rm rel\,LMC}$, in each bandpass.  These best-fit relationships are within the uncertainties for the best fits when no clipping is applied, because about half of the outliers are above and half are below the $3\sigma$ distribution of candidates. 

\begin{deluxetable}{lcccc}
\renewcommand{\arraystretch}{2}
\tablecolumns{5}
\tablewidth{0pt}
\tablecaption{Period-Luminosity Relationships: $m = \alpha \log P + \beta$}
\tablehead{
\colhead{Band} &
\colhead{$\alpha$} &
\colhead{$\beta$} &
\colhead{$\sigma$} &
\colhead{$\mu_{\rm rel\,LMC}$} 
\\
\colhead{} &
\colhead{} &
\colhead{} &
\colhead{(dex)} &
\colhead{(mag)}
}
\startdata
F555W & $-2.76$ & $31.464^{+0.035}_{-0.033}$  & $0.099^{+0.018}_{-0.015}$  & $13.826^{+0.035}_{-0.033}$ \\
F814W & $-2.96$ & $30.371^{+0.026}_{-0.026}$  & $0.050^{+0.010}_{-0.008}$  & $13.517^{+0.026}_{-0.026}$ \\
W$_I$ & $-3.31$ & $29.084^{+0.032}_{-0.033}$  & $0.068^{+0.016}_{-0.013}$  & $13.149^{+0.032}_{-0.033}$ \\
{\textbf W$_I$ ({\textit P>21}\,d)} & $-3.31$ & $29.135^{+0.031}_{-0.031}$ &  $0.051^{+0.014}_{-0.011}$ & $13.200^{+0.031}_{-0.031}$
\label{tab:pl}
\enddata

\tablecomments{$\alpha$ was held fixed at the values used by \citet{riess19} in their analysis of LMC Cepheids.}
\end{deluxetable}

\section{Discussion}

It is clear based on the results listed in Table~\ref{tab:pl} that the distance modulus relative to the LMC based on fits to the F555W and F814W magnitudes differ substantially from that derived from the Wesenheit magnitudes.  Given the equatorial coordinates of NGC\,6814, we are viewing the galaxy through the disk of the Milky Way and so substantial extinction along the line of sight is expected.  Thus the F555W photometry, and to a lesser extent the F814W photometry, will underestimate the magnitudes of the Cepheid candidates and overestimate the distance modulus.  The Wesenheit magnitudes, by definition, are not susceptible to this same effect. Based on the \citet{schlafly11} recalibration of the \citet{schlegel98} dust map of the Milky Way, we can estimate $A_V=0.509$ and $A_I=0.279$ along the line of sight to NGC\,6814 with a nominal uncertainty of 0.040\,mag.  Correcting for extinction, the F555W and F814W photometry then predict distance moduli relative to the LMC of $13.317^{+0.053}_{-0.052}$\,mag and $13.238^{+0.048}_{-0.048}$\,mag, respectively.  These values are consistent with each other and with the value of $\mu_{\rm rel\,LMC}$ predicted from the Wesenheit magnitudes.

A potential complication is that the Cepheid period-luminosity relationships will generally suffer from incompleteness at the low-luminosity, short-period end due to the flux-limited nature of the observations.   We therefore investigated the effects of incompleteness by cutting the sample of candidates at various minimum period values and redetermining the best-fit $W_I$ relationship.  The effect on $\mu_{\rm rel\,LMC}$ as a function of minimum period is displayed in Figure~\ref{fig:minp}.  We detect clear signatures of incompleteness for $P_{min} \lesssim 21$\,days (as denoted by the dotted vertical line in Figure~\ref{fig:minp}).    This is similar to the findings of \citet{riess11} who determined that incompleteness generally sets in for Cepheids with S/N<10, which we previously estimated to occur around $P \approx 20$\,days for NGC\,6814.

We therefore report the best-fit $W_I$ relationship for the $\sim80$\% of the sample with $P>21$\,days as our best determination of the distance modulus relative to the LMC for NGC\,6814, $\mu_{\rm rel\,LMC}=13.200^{+0.031}_{-0.031}$\,mag.  If we separately consider the candidates that are $<1\farcm0$ and $>1\farcm0$ from the center of the galaxy (roughly 50\% of the candidates in each bin), we find distance moduli that are completely consistent with the result using all of the candidates, demonstrating that there does not appear to be a bias in crowding or extinction as a function of radius across this nearly face-on galaxy. 

The distance modulus of the LMC itself has been an elusive target for many years (see, e.g., \citealt{freedman10,walker12}).  Recently \citet{pietrzynski19} determined a distance modulus to the LMC that is precise to 1\%,  $18.477 \pm 0.026$\,mag, based on the geometric distances determined for 20 detached eclipsing binary systems. Adopting the \citet{pietrzynski19} distance modulus to the LMC, we find $m-M=31.677^{+0.041}_{-0.041}$\,mag which corresponds to a distance of $D=21.65\pm0.41$\,Mpc to NGC\,6814.  

We find a similar distance if we instead use the \citet{soszynski16} relationships for fundamental pulsating Cepheids in the LMC from OGLE.  In this case, we convert our F555W and F814W magnitudes to $V-$ and $I-$band magnitudes with small color terms of $c_V=-0.086$\,mag and $c_I=-0.003$\,mag determined from synthetic photometry of F, G, and K giant stars by \citet{sahu14}.   \citet{soszynski16} defined their Wesenheit magnitudes with a slightly different extinction law, adopting $R=1.55$.  With these minor changes, and again fitting only the candidates with $P>21$\,days, we find $\mu_{\rm rel\,LMC}=12.972^{+0.046}_{-0.048}$\,mag.  Again adopting the \citet{pietrzynski19} distance modulus for the LMC, we find $m-M=31.449^{+0.053}_{-0.055}$\,mag and $D=19.49\pm0.48$\,Mpc.  The two results are consistent within 3$\sigma$, but we prefer the distance based on the \citet{riess19} calibration of Cepheids within the LMC because it avoids the uncertainties and potential biases inherent in color corrections between different filters.

The effect of metallicity on Cepheid-based distances is an ongoing area of study.  The LMC is well-known to have a lower average metallicity than the Milky Way and other massive galaxies.  \citet{sakai04} and several other studies find that Cepheids in galaxies with metal-rich HII regions are systematically brighter and thus underestimate the distance modulus compared to that obtained from the tip of the red giant branch method.  On the other hand, \citet{romaniello08} find that metal-rich Cepheids are systematically {\it fainter} when using Fe abundances determined from direct spectroscopy of individual Cepheids in the Galaxy, LMC, and SMC.  As it is not possible to obtain direct spectroscopy of individual Cepheids in most other galaxies of interest, the tension in these results has not yet been resolved. In the case of NGC\,6814, little information on the metallicity of the galaxy currently exists.  \citet{ryder05} measured the Lick indices Mg$_2$ and Fe\,5270 as a function of radius across NGC\,6814 and they find a shallow dependence on radius outside the region influenced by the AGN, with an average abundance that is approximately solar.  Based on this, we would expect that any metallicity correction to the distance modulus for the Cepheids in NGC\,6814 would be small, on the order of $\sim0.05$\,mag following the \citet{sakai04} method, and is therefore on the order of our formal uncertainties.

With an accurate distance measurement now in hand for NGC\,6814, we can determine the magnitude of peculiar velocities affecting its apparent redshift.  \citet{robinson19} recently reported measurements of the HI 21\,cm emission line from Green Bank Telescope observations of NGC\,6814. They determined $v_{rec} = 1562$\,km\,s$^{-1}$, which predicts $D=21.1$\,Mpc when adopting $H_0 = 74$\,km\,s$^{-1}$\,Mpc$^{-1}$ \citep{riess19} or $D=23.2$\,Mpc when adopting $H_0 = 67.4$\,km\,s$^{-1}$\,Mpc$^{-1}$ \citep{planck18}.  Our Cepheid-based distance is consistent and implies that peculiar velocities on the order of $\lesssim100$\,km\,s$^{-1}$ are affecting the recessional velocity of NGC\,6814.  

NGC\,6814 is an isolated galaxy, so it is perhaps unsurprising that it is not experiencing any particularly strong peculiar velocities. Nevertheless, peculiar velocities have been measured to be as large as $500$\,km\,s$^{-1}$ \citep{tully13} for galaxies within the local universe, and thus distance measurements that do not rely on the redshift of the galaxy are critical for determining accurate luminosities and dynamical masses. The Cepheid-based distance derived here for NGC\,6814 will ensure that our efforts to constrain a black hole mass based on stellar dynamical modeling will not suffer from the uncertainties introduced by an inaccurate distance.

\section{Summary}
We have presented a {\it Hubble Space Telescope} monitoring program to detect and characterize Cepheids in the nearby Seyfert galaxy NGC\,6814.  Based on the analysis of Cepheid candidates detected above the incompleteness limit, we derive a distance modulus relative to the LMC of $\mu_{\rm rel\,LMC}=13.200^{+0.031}_{-0.031}$\,mag. Adopting the \citet{pietrzynski19} constraint on the distance modulus to the LMC, we find $m-M=31.677^{+0.041}_{-0.041}$\,mag which gives a distance of $21.65\pm0.41$\,Mpc to NGC\,6814.  The Cepheid-based distance measurement is consistent with the distance predicted by the redshift of the HI 21\,cm emission line, implying that this relatively isolated galaxy is experiencing peculiar velocities of $\lesssim 100$\,km\,s$^{-1}$.

\acknowledgements

We thank the referee for helpful comments that improved the presentation of this work. MCB gratefully acknowledges support from the NSF through CAREER grant AST-1253702. We are grateful for support of this work through grant HST GO-12961 from the Space Telescope Science Institute, which is operated by the Association of Universities for Research in Astronomy, Inc., under NASA contract NAS5-26555. This research has made use of the NASA/IPAC Extragalactic Database (NED) which is operated by the Jet Propulsion Laboratory, California Institute of Technology, under contract with the National Aeronautics and Space Administration and the SIMBAD database, operated at CDS, Strasbourg, France.


\clearpage

\LongTables
\begin{deluxetable*}{lccccccc}
\tablecolumns{8}
\tablewidth{0pt}
\tablecaption{Best-Fit Properties of Cepheid Candidates}
\tablehead{
\colhead{\#} &
\colhead{$\alpha$} &
\colhead{$\delta$} &
\colhead{$m_{F555W}$} &
\colhead{$m_{F814W}$} &
\colhead{$P$} &
\colhead{phase\tablenotemark{a}} &
\colhead{$\chi^2/\nu$}
\\
\colhead{} &
\colhead{(hh:mm:ss)} &
\colhead{(dd:mm:ss)} &
\colhead{(mag)} &
\colhead{(mag)} &
\colhead{(days)} &
\colhead{(days)} &
\colhead{}
}
\startdata
1   & $19:42:43.690$  &   $-10:20:10.475$ &  $25.83 \pm 0.03$ &  $24.58 \pm 0.04$ &  $83.57 \pm 9.52$ &  $16.82 \pm 2.53$ &  1.87  \\ %
2   & $19:42:42.332$  &   $-10:19:40.200$ &  $26.32 \pm 0.02$ &  $24.74 \pm 0.03$ &  $82.47 \pm 7.42$ &  $43.60 \pm 1.26$ &  0.88  \\ %
3   & $19:42:38.881$  &   $-10:19:30.575$ &  $25.82 \pm 0.02$ &  $24.59 \pm 0.03$ &  $78.72 \pm 6.23$ &  $10.98 \pm 2.49$ &  0.44  \\ %
4   & $19:42:41.872$  &   $-10:19:54.003$ &  $25.62 \pm 0.03$ &  $24.61 \pm 0.05$ &  $78.18 \pm 0.24$ &  $55.18 \pm 2.33$ &  3.90  \\ %
5   & $19:42:45.903$  &   $-10:20:02.840$ &  $27.44 \pm 0.03$ &  $25.66 \pm 0.04$ &  $75.60 \pm 9.05$ &  $48.10 \pm 2.09$ &  0.76  \\ %
6   & $19:42:40.864$  &   $-10:18:20.589$ &  $26.50 \pm 0.04$ &  $24.99 \pm 0.05$ &  $69.03 \pm 6.38$ &  $ 5.04 \pm 5.60$ &  2.27  \\ %
7   & $19:42:44.737$  &   $-10:19:32.311$ &  $26.27 \pm 0.03$ &  $24.86 \pm 0.05$ &  $65.75 \pm 3.74$ &  $44.43 \pm 0.92$ &  1.97  \\ %
8   & $19:42:39.046$  &   $-10:20:33.986$ &  $26.69 \pm 0.03$ &  $25.22 \pm 0.05$ &  $56.76 \pm 3.15$ &  $40.87 \pm 0.96$ &  2.11  \\ %
9   & $19:42:37.890$  &   $-10:19:02.948$ &  $26.94 \pm 0.03$ &  $25.61 \pm 0.05$ &  $54.23 \pm 3.43$ &  $48.13 \pm 0.87$ &  1.12  \\ %
10  & $19:42:42.298$  &   $-10:18:54.908$ &  $26.67 \pm 0.02$ &  $25.46 \pm 0.05$ &  $52.15 \pm 2.41$ &  $47.75 \pm 0.56$ &  0.97  \\ %
11  & $19:42:40.062$  &   $-10:20:02.234$ &  $26.86 \pm 0.04$ &  $25.48 \pm 0.08$ &  $51.88 \pm 1.88$ &  $17.92 \pm 1.62$ &  1.99  \\ %
12  & $19:42:41.262$  &   $-10:18:47.769$ &  $26.70 \pm 0.03$ &  $25.42 \pm 0.05$ &  $51.32 \pm 2.03$ &  $ 4.77 \pm 1.94$ &  0.98  \\ %
13  & $19:42:38.180$  &   $-10:18:46.790$ &  $26.46 \pm 0.02$ &  $25.12 \pm 0.04$ &  $51.03 \pm 1.39$ &  $22.45 \pm 1.22$ &  1.25  \\ %
14  & $19:42:36.214$  &   $-10:19:58.432$ &  $26.80 \pm 0.04$ &  $25.52 \pm 0.06$ &  $48.40 \pm 2.30$ &  $ 3.66 \pm 2.53$ &  2.17  \\ %
15  & $19:42:44.504$  &   $-10:19:23.579$ &  $26.90 \pm 0.04$ &  $25.61 \pm 0.07$ &  $47.14 \pm 2.13$ &  $29.58 \pm 1.13$ &  2.12  \\ %
16  & $19:42:36.965$  &   $-10:19:17.924$ &  $27.54 \pm 0.04$ &  $26.02 \pm 0.07$ &  $47.03 \pm 1.54$ &  $24.24 \pm 1.02$ &  1.08  \\ %
17  & $19:42:43.571$  &   $-10:20:09.477$ &  $26.69 \pm 0.03$ &  $25.35 \pm 0.05$ &  $45.87 \pm 1.70$ &  $38.11 \pm 0.58$ &  1.65  \\ %
18  & $19:42:39.356$  &   $-10:18:33.040$ &  $27.06 \pm 0.03$ &  $25.48 \pm 0.05$ &  $43.88 \pm 0.82$ &  $32.06 \pm 0.41$ &  1.08  \\ %
19  & $19:42:38.851$  &   $-10:20:29.338$ &  $26.96 \pm 0.03$ &  $25.67 \pm 0.05$ &  $43.45 \pm 2.60$ &  $19.02 \pm 2.19$ &  1.01  \\ %
20  & $19:42:44.267$  &   $-10:19:32.738$ &  $26.87 \pm 0.03$ &  $25.55 \pm 0.07$ &  $43.26 \pm 2.52$ &  $34.89 \pm 0.41$ &  1.34  \\ %
21  & $19:42:38.119$  &   $-10:19:47.794$ &  $27.02 \pm 0.04$ &  $25.40 \pm 0.06$ &  $43.20 \pm 0.74$ &  $15.55 \pm 0.54$ &  2.04  \\ %
22  & $19:42:36.014$  &   $-10:19:42.052$ &  $26.81 \pm 0.03$ &  $25.47 \pm 0.05$ &  $42.54 \pm 0.96$ &  $14.20 \pm 0.50$ &  1.83  \\ %
23  & $19:42:42.917$  &   $-10:19:54.123$ &  $27.67 \pm 0.06$ &  $26.13 \pm 0.09$ &  $42.24 \pm 2.16$ &  $27.74 \pm 1.50$ &  1.24  \\ %
24  & $19:42:38.886$  &   $-10:19:51.284$ &  $27.45 \pm 0.05$ &  $25.79 \pm 0.07$ &  $41.61 \pm 1.18$ &  $12.93 \pm 0.90$ &  1.28  \\ %
25  & $19:42:41.796$  &   $-10:19:02.093$ &  $27.40 \pm 0.04$ &  $25.87 \pm 0.06$ &  $40.73 \pm 2.97$ &  $22.12 \pm 1.85$ &  0.81  \\ %
26  & $19:42:38.175$  &   $-10:18:57.382$ &  $27.55 \pm 0.07$ &  $25.77 \pm 0.10$ &  $39.09 \pm 1.62$ &  $33.94 \pm 1.29$ &  3.00  \\ %
27  & $19:42:44.278$  &   $-10:19:15.848$ &  $26.69 \pm 0.02$ &  $25.41 \pm 0.04$ &  $38.89 \pm 1.25$ &  $ 1.90 \pm 1.30$ &  0.83  \\ %
28  & $19:42:43.026$  &   $-10:19:06.632$ &  $26.81 \pm 0.06$ &  $25.69 \pm 0.13$ &  $37.73 \pm 2.15$ &  $ 7.09 \pm 2.05$ &  3.85  \\ %
29  & $19:42:38.589$  &   $-10:20:33.130$ &  $27.67 \pm 0.04$ &  $26.19 \pm 0.07$ &  $37.28 \pm 0.71$ &  $35.10 \pm 0.35$ &  0.94  \\ %
30  & $19:42:43.464$  &   $-10:19:45.077$ &  $27.08 \pm 0.05$ &  $25.93 \pm 0.11$ &  $37.18 \pm 1.03$ &  $14.71 \pm 0.64$ &  3.12  \\ %
31  & $19:42:42.726$  &   $-10:19:38.815$ &  $27.18 \pm 0.05$ &  $25.82 \pm 0.09$ &  $36.55 \pm 1.00$ &  $23.19 \pm 0.68$ &  1.57  \\ %
32  & $19:42:39.062$  &   $-10:18:34.258$ &  $27.39 \pm 0.04$ &  $25.93 \pm 0.09$ &  $36.45 \pm 2.74$ &  $10.93 \pm 2.71$ &  1.49  \\ %
33  & $19:42:38.427$  &   $-10:19:24.693$ &  $27.02 \pm 0.02$ &  $25.85 \pm 0.05$ &  $36.37 \pm 0.40$ &  $31.63 \pm 0.28$ &  0.64  \\ %
34  & $19:42:38.248$  &   $-10:20:30.004$ &  $27.50 \pm 0.05$ &  $26.20 \pm 0.08$ &  $36.30 \pm 1.89$ &  $26.66 \pm 0.85$ &  1.17  \\ %
35  & $19:42:40.305$  &   $-10:18:41.179$ &  $27.46 \pm 0.04$ &  $25.96 \pm 0.07$ &  $35.60 \pm 0.98$ &  $25.51 \pm 0.76$ &  1.15  \\ %
36  & $19:42:40.721$  &   $-10:18:14.427$ &  $27.17 \pm 0.06$ &  $25.88 \pm 0.10$ &  $34.82 \pm 3.06$ &  $20.71 \pm 2.88$ &  3.41  \\ %
37  & $19:42:37.179$  &   $-10:20:26.045$ &  $27.31 \pm 0.03$ &  $26.03 \pm 0.06$ &  $34.56 \pm 0.79$ &  $ 4.05 \pm 1.44$ &  0.92  \\ %
38  & $19:42:42.355$  &   $-10:19:42.509$ &  $26.72 \pm 0.06$ &  $25.43 \pm 0.09$ &  $34.31 \pm 1.10$ &  $12.80 \pm 0.79$ &  3.98  \\ %
39  & $19:42:44.327$  &   $-10:18:45.457$ &  $27.01 \pm 0.03$ &  $25.78 \pm 0.06$ &  $34.23 \pm 1.32$ &  $ 8.05 \pm 1.32$ &  1.53  \\ %
40  & $19:42:36.797$  &   $-10:19:27.006$ &  $27.16 \pm 0.07$ &  $25.89 \pm 0.09$ &  $34.15 \pm 1.74$ &  $15.59 \pm 0.93$ &  2.02  \\ %
41  & $19:42:39.336$  &   $-10:19:59.592$ &  $26.99 \pm 0.04$ &  $26.10 \pm 0.11$ &  $33.13 \pm 0.75$ &  $14.58 \pm 0.40$ &  1.43  \\ %
42  & $19:42:43.183$  &   $-10:19:50.198$ &  $27.22 \pm 0.04$ &  $25.97 \pm 0.09$ &  $32.89 \pm 0.55$ &  $ 8.58 \pm 0.84$ &  1.76  \\ %
43  & $19:42:37.285$  &   $-10:19:57.179$ &  $27.25 \pm 0.04$ &  $25.99 \pm 0.07$ &  $32.46 \pm 1.68$ &  $21.71 \pm 1.61$ &  1.42  \\ %
44  & $19:42:42.277$  &   $-10:20:13.806$ &  $27.49 \pm 0.05$ &  $26.01 \pm 0.08$ &  $31.35 \pm 0.79$ &  $ 2.91 \pm 1.07$ &  1.38  \\ %
45  & $19:42:43.074$  &   $-10:20:50.872$ &  $26.90 \pm 0.05$ &  $25.73 \pm 0.10$ &  $30.95 \pm 0.79$ &  $15.26 \pm 0.71$ &  3.41  \\ %
46  & $19:42:43.663$  &   $-10:20:12.710$ &  $27.44 \pm 0.14$ &  $26.14 \pm 0.11$ &  $30.63 \pm 1.74$ &  $20.32 \pm 3.53$ &  1.39  \\ %
47  & $19:42:38.208$  &   $-10:19:07.581$ &  $27.35 \pm 0.05$ &  $26.23 \pm 0.10$ &  $29.88 \pm 1.29$ &  $ 2.30 \pm 1.43$ &  1.16  \\ %
48  & $19:42:40.114$  &   $-10:18:42.237$ &  $27.28 \pm 0.16$ &  $26.02 \pm 0.12$ &  $29.66 \pm 1.68$ &  $19.97 \pm 3.95$ &  1.13  \\ %
49  & $19:42:43.348$  &   $-10:20:28.990$ &  $26.99 \pm 0.03$ &  $25.95 \pm 0.08$ &  $29.41 \pm 0.76$ &  $ 7.54 \pm 1.45$ &  1.46  \\ %
50  & $19:42:40.047$  &   $-10:18:42.167$ &  $27.19 \pm 0.03$ &  $25.93 \pm 0.05$ &  $29.31 \pm 1.12$ &  $ 6.46 \pm 1.16$ &  0.87  \\ %
51  & $19:42:39.014$  &   $-10:18:09.166$ &  $27.57 \pm 0.04$ &  $25.96 \pm 0.06$ &  $28.35 \pm 0.39$ &  $13.64 \pm 0.30$ &  1.01  \\ %
52  & $19:42:41.158$  &   $-10:20:11.595$ &  $27.45 \pm 0.05$ &  $26.03 \pm 0.09$ &  $28.18 \pm 0.88$ &  $15.29 \pm 0.85$ &  1.73  \\ %
53  & $19:42:43.338$  &   $-10:19:50.490$ &  $27.19 \pm 0.05$ &  $26.02 \pm 0.12$ &  $28.15 \pm 0.66$ &  $11.21 \pm 0.99$ &  3.36  \\ %
54  & $19:42:39.116$  &   $-10:18:37.673$ &  $27.55 \pm 0.04$ &  $26.19 \pm 0.09$ &  $28.00 \pm 0.93$ &  $ 4.64 \pm 1.67$ &  1.15  \\ %
55  & $19:42:43.249$  &   $-10:19:52.735$ &  $27.05 \pm 0.05$ &  $25.96 \pm 0.12$ &  $27.70 \pm 0.45$ &  $17.29 \pm 0.68$ &  3.26  \\ %
56  & $19:42:35.703$  &   $-10:19:56.981$ &  $27.48 \pm 0.06$ &  $26.16 \pm 0.12$ &  $27.28 \pm 2.34$ &  $17.01 \pm 2.35$ &  3.06  \\ %
57  & $19:42:43.183$  &   $-10:20:37.411$ &  $27.29 \pm 0.14$ &  $26.03 \pm 0.11$ &  $26.66 \pm 1.02$ &  $11.59 \pm 3.12$ &  2.36  \\ %
58  & $19:42:36.046$  &   $-10:19:23.893$ &  $27.65 \pm 0.04$ &  $26.23 \pm 0.07$ &  $25.20 \pm 0.78$ &  $ 8.13 \pm 1.50$ &  0.94  \\ %
59  & $19:42:42.572$  &   $-10:21:04.353$ &  $27.90 \pm 0.06$ &  $26.58 \pm 0.13$ &  $24.36 \pm 1.24$ &  $19.57 \pm 1.35$ &  1.83  \\ %
60  & $19:42:43.244$  &   $-10:20:10.482$ &  $27.33 \pm 0.06$ &  $26.26 \pm 0.12$ &  $24.08 \pm 0.91$ &  $16.58 \pm 1.05$ &  2.11  \\ %
61  & $19:42:40.388$  &   $-10:18:10.924$ &  $27.32 \pm 0.05$ &  $26.10 \pm 0.12$ &  $24.00 \pm 1.05$ &  $12.19 \pm 1.93$ &  2.52  \\ %
62  & $19:42:42.858$  &   $-10:20:07.863$ &  $27.80 \pm 0.06$ &  $26.29 \pm 0.09$ &  $23.79 \pm 0.59$ &  $ 6.42 \pm 0.87$ &  1.21  \\ %
63  & $19:42:41.165$  &   $-10:18:09.605$ &  $27.44 \pm 0.05$ &  $26.02 \pm 0.10$ &  $23.61 \pm 0.52$ &  $ 6.94 \pm 0.77$ &  1.83  \\ %
64  & $19:42:39.344$  &   $-10:17:58.410$ &  $28.08 \pm 0.07$ &  $26.82 \pm 0.14$ &  $23.35 \pm 0.31$ &  $13.28 \pm 0.36$ &  1.05  \\ %
65  & $19:42:36.862$  &   $-10:19:20.011$ &  $27.54 \pm 0.05$ &  $26.36 \pm 0.10$ &  $23.29 \pm 1.31$ &  $22.59 \pm 1.33$ &  1.39  \\ %
66  & $19:42:44.142$  &   $-10:19:35.823$ &  $27.37 \pm 0.07$ &  $26.32 \pm 0.15$ &  $23.15 \pm 0.71$ &  $21.66 \pm 0.97$ &  2.98  \\ %
67  & $19:42:35.013$  &   $-10:19:47.240$ &  $27.73 \pm 0.05$ &  $26.44 \pm 0.08$ &  $23.01 \pm 0.64$ &  $ 5.74 \pm 1.64$ &  0.80  \\ %
68  & $19:42:38.166$  &   $-10:20:23.127$ &  $27.47 \pm 0.04$ &  $26.51 \pm 0.09$ &  $22.76 \pm 0.26$ &  $12.83 \pm 0.35$ &  0.99  \\ %
69  & $19:42:42.506$  &   $-10:19:15.705$ &  $27.27 \pm 0.06$ &  $26.20 \pm 0.10$ &  $22.63 \pm 0.70$ &  $ 5.41 \pm 1.95$ &  0.94  \\ %
70  & $19:42:44.176$  &   $-10:20:41.015$ &  $27.94 \pm 0.07$ &  $26.47 \pm 0.12$ &  $22.13 \pm 0.89$ &  $ 9.51 \pm 1.67$ &  1.88  \\ %
71  & $19:42:40.890$  &   $-10:20:52.821$ &  $27.26 \pm 0.06$ &  $26.14 \pm 0.12$ &  $21.88 \pm 1.11$ &  $16.53 \pm 2.11$ &  2.76  \\ %
72  & $19:42:41.556$  &   $-10:20:06.796$ &  $27.44 \pm 0.03$ &  $26.19 \pm 0.07$ &  $21.65 \pm 0.56$ &  $20.49 \pm 0.61$ &  0.82  \\ %
73  & $19:42:38.283$  &   $-10:20:06.382$ &  $27.78 \pm 0.06$ &  $26.31 \pm 0.13$ &  $21.33 \pm 1.00$ &  $17.59 \pm 1.72$ &  2.26  \\ %
74  & $19:42:44.644$  &   $-10:19:27.905$ &  $27.70 \pm 0.03$ &  $26.55 \pm 0.08$ &  $21.19 \pm 0.53$ &  $20.09 \pm 0.64$ &  0.59  \\ %
75  & $19:42:44.673$  &   $-10:20:04.569$ &  $28.05 \pm 0.07$ &  $26.70 \pm 0.13$ &  $21.14 \pm 0.48$ &  $13.64 \pm 0.44$ &  1.30  \\ %
76  & $19:42:37.151$  &   $-10:20:19.046$ &  $27.85 \pm 0.05$ &  $26.61 \pm 0.11$ &  $21.08 \pm 0.75$ &  $ 1.12 \pm 1.57$ &  1.20  \\ %
77  & $19:42:42.540$  &   $-10:19:03.332$ &  $27.27 \pm 0.05$ &  $25.88 \pm 0.09$ &  $20.77 \pm 0.68$ &  $ 0.54 \pm 1.43$ &  1.81  \\ %
78  & $19:42:40.304$  &   $-10:18:05.329$ &  $27.30 \pm 0.05$ &  $26.15 \pm 0.10$ &  $20.21 \pm 0.64$ &  $ 1.79 \pm 1.34$ &  1.54  \\ %
79  & $19:42:43.802$  &   $-10:18:42.254$ &  $27.55 \pm 0.05$ &  $26.12 \pm 0.08$ &  $20.04 \pm 0.60$ &  $10.02 \pm 0.75$ &  1.58  \\ %
80  & $19:42:40.094$  &   $-10:20:35.717$ &  $27.76 \pm 0.05$ &  $26.68 \pm 0.13$ &  $19.88 \pm 0.90$ &  $18.78 \pm 1.63$ &  1.10  \\ %
81  & $19:42:37.426$  &   $-10:20:11.307$ &  $27.74 \pm 0.07$ &  $26.31 \pm 0.13$ &  $19.17 \pm 0.54$ &  $14.87 \pm 0.97$ &  2.33  \\ %
82  & $19:42:43.945$  &   $-10:18:35.666$ &  $27.62 \pm 0.05$ &  $26.28 \pm 0.09$ &  $19.08 \pm 0.55$ &  $10.69 \pm 0.85$ &  1.37  \\ %
83  & $19:42:38.351$  &   $-10:19:01.362$ &  $27.40 \pm 0.06$ &  $26.01 \pm 0.10$ &  $18.82 \pm 0.59$ &  $ 4.48 \pm 0.99$ &  1.46  \\ %
84  & $19:42:35.440$  &   $-10:19:41.352$ &  $27.59 \pm 0.04$ &  $26.33 \pm 0.09$ &  $18.48 \pm 0.27$ &  $12.46 \pm 0.49$ &  1.00  \\ %
85  & $19:42:39.791$  &   $-10:18:00.931$ &  $27.85 \pm 0.04$ &  $26.66 \pm 0.10$ &  $18.42 \pm 0.45$ &  $ 6.44 \pm 0.91$ &  0.86  \\ %
86  & $19:42:41.842$  &   $-10:19:38.874$ &  $27.73 \pm 0.05$ &  $26.45 \pm 0.09$ &  $17.88 \pm 0.36$ &  $ 6.39 \pm 0.59$ &  0.82  \\ %
87  & $19:42:37.728$  &   $-10:20:33.964$ &  $27.47 \pm 0.05$ &  $26.39 \pm 0.09$ &  $17.14 \pm 0.70$ &  $ 3.31 \pm 1.48$ &  1.19  \\ %
88  & $19:42:36.029$  &   $-10:19:53.987$ &  $27.97 \pm 0.07$ &  $26.51 \pm 0.12$ &  $16.91 \pm 0.65$ &  $ 8.84 \pm 1.67$ &  1.54  \\ %
89  & $19:42:46.808$  &   $-10:18:53.823$ &  $27.15 \pm 0.04$ &  $26.35 \pm 0.08$ &  $16.35 \pm 0.35$ &  $ 3.53 \pm 1.04$ &  0.83  \\ %
90  & $19:42:34.940$  &   $-10:19:59.494$ &  $28.17 \pm 0.06$ &  $26.88 \pm 0.11$ &  $13.10 \pm 0.37$ &  $ 2.71 \pm 1.13$ &  0.96  \\ %
\label{tab:cands}
\enddata 
\tablenotetext{a}{The phase of the light curve is relative to the date of the first visit.}

\end{deluxetable*}

\clearpage


\end{document}